\documentclass[conference]{IEEEtran}
\IEEEoverridecommandlockouts
\usepackage{cite}
\usepackage{amsmath,amssymb,amsfonts}
\usepackage{algorithmic}
\usepackage{graphicx}
\usepackage{textcomp}
\usepackage{xcolor}
\usepackage{hyperref}
\def\BibTeX{{\rm B\kern-.05em{\sc i\kern-.025em b}\kern-.08em
    T\kern-.1667em\lower.7ex\hbox{E}\kern-.125emX}}
\begin{document}

\title{Re-Imagining Performance Reviews: Automated Dashboards for Continuous Visibility of Engineers’ Performance\\

}

\author{\IEEEauthorblockN{Fatma Meawad}
\IEEEauthorblockA{\textit{Robusta} \\
Cairo, Egypt \\
fatma.meawad@robustastudio.com}
}

\maketitle


\begin{abstract}

Traditional methods for performance appraisal are not suitable for agile fast-paced software companies. This has been a realization in the software industry since the early adoption of agile methodologies. Nonetheless, software companies are still struggling to find a practical solution that fits the highly dynamic nature of their teams. In particular, high-growth companies, scaleups, need to be creative at how they approach performance appraisals. 

In this talk, we introduce automated review dashboards to support a seamless appraisal process and continuous visibility of software engineers' performance.  The proposed dashboards leverage structure from existing growth frameworks while automating the aggregation of relevant qualitative and quantitative performance metrics. 

We reflect on our experience using the dashboards at Robusta, a medium-sized software scaleup company. The dashboards enabled a team of four managers to provide timely feedback to 56 engineers with a base for continuous visibility. We explore the design of the dashboards, the customizable metrics and the overall review experience from the perspectives of different stakeholders. We conclude with the lessons learned and practical advice for scaleups facing the same challenge.

\end{abstract}


\section{The Key points of the talk}

There is a strong need to rethink the way performance appraisals are done for software engineers, particularly in scale-up agile companies. Traditional methods for performance reviews and appraisals require a significant investment from a team’s capacity which does not suit the dynamic nature of agile teams \cite{cappelli2016performance, alanatlas,rejab2018transition, al2018individuals,goler2016let}. Such investment is wasted if the feedback is not timely and relevant to the team’s working conditions.  In other words, performance reviews need to be agile, by allowing frequent relevant feedback and continuous visibility of an engineers’ growth. At Robusta\footnote{https://robustastudio.com/}, we attempted to automate the appraisal process while maintaining its value through performance dashboards. This talk includes the following key points from our experience:

\subsection{Background and Motivation}

Robusta is a software house providing design, development and support of custom products for its clients. Robusta handles large and complex projects for multiple clients simultaneously in Egypt and Germany. The engineering function officially uses Scrum or Kanban; however, the development team, including managers, are required to adapt the process to the unpredictable demands on delivery while maintaining the agile values. 

One realization from Google's project Oxygen \cite{garvin2013google}, is that software engineers hate micromanagement but are keen to get frequent feedback about their growth from their managers. Therefore, even with the intense time-pressure and uncertainty in delivery, mentoring and leveling up the team are among the top priorities for a scaleup such as Robusta. With a strong competition in the market for the best engineers, maintaining a healthy growth culture is key to the success of the company.


\subsection{Design \& Implementation} 

Identifying  how to evaluate agile software engineers is a difficult problem \cite{alnaji2015performance}. Some work in literature explore this topic \cite{pack2010evaluating, rejab2018transition, turley1995competencies}, but none addresses the case for teams with dynamic allocations and unpredictable project conditions. With a focus on growth, we came up with 
an initial dashboard design that is customizable, enables automation and accommodates exceptional cases.




There are three main aspects to our design: the high level structure, the competencies and the metrics. The high level structure is inspired by several growth framework published by technology companies such as Medium \cite{medium}, Expert 360 \cite{expert360} and Spotify \cite{spotify}. The structure includes six main topics: (a) efficiency \& quality, (b) technical competencies,(c) recruitment, (d) leadership, (e) learning \& development and (f) community. These six topics act as goals that cover all aspects of growth for all the engineers in different roles. For each topic, engineers are rated according to their impact level: (a) individual, (b) across teams, (c) across the organization or (d) industry. 

Under each topic, relevant competencies are identified.
In addition, multiple metrics are extracted from projects' data to describe how an engineer meets a certain competency. Relying on multiple metrics for the same competency helps in clarifying cases for teams working under different conditions.

The selected metrics can be (a) quantitative, directly  calculated from  project's data  or (b) qualitative, aggregated from peer's feedback for the engineer. The metrics can change based on engineers' feedback or to reflect any change in the corresponding working conditions.








The review dashboard is currently an internal product at Robusta in its alpha release phase. The chosen metrics are aggregated from Jira, Gitlab and peer feedback responses from Google forms. The whole dashboard is presented on google sheets using Google's Apps Script. 
In the talk, we will present a demo of the dashboards and discuss the rationale for choosing the different metrics and competencies.


\subsection{Results and Lessons Learned}


Not only did the dashboards accelerate the performance review process, but also they had an unexpected impact on the team's morale and perception of the appraisal process. Data enthusiasts were excited about the data analysis involved and process enthusiasts were excited about alignment and organization. In the meantime, supervisors were relieved from the preparation overhead and the threat of subjectivity in such process. Furthermore, the structure and metrics introduced in the dashboards enforced consistency and alignment about process and expectations in the engineering function. Overall, bringing such attention to the alignment on quality, process, learning \& development, community and leadership has brought about a new spirit in the engineering function that appreciates growth as a whole, not just meeting delivery needs.

The current version of the dashboards is only a start. It helped us attract valuable feedback about additional metrics and additional sources for data collection. There were some concerns about the validity of the metrics and if they truly reflect the exceptional circumstances for different projects. The dashboard can accommodate the proposed changes to the metrics to fit under the existing competencies. We look forward to sharing lessons learned about the results of the experience with practitioners and academics in the field. 


\section*{Speaker's Biography}
Fatma Meawad is a computer scientist and the Director of Engineering
at Robusta. Fatma holds a PhD in Computer Science from the University of South Wales, a Masters in Advanced Computing from the University Of Birmingham and a BSc in Computer Science from the American University in Cairo. Fatma has 10 years of experience teaching Computer Science at different Universities in the UK, Egypt, and Singapore. As an academic, Fatma is well known for establishing strong engagements with industry to
enhance the quality of education, especially in the area of software
engineering. Fatma’s latest research achievements are in mobile health
and personal informatics. As an associate professor at Singapore
Institute of Technology (SIT), Fatma led multiple research grants in
leveraging multimodal sensing for developing context-aware
psychologically-sound interventions for chronic pain patients.

\section*{Acknowledgment}
I would like to thank Ahmed Alfy, the Chief Technology Officer at Robusta for the insightful discussions and for providing the gitlab data.

\bibliographystyle{./bibliography/IEEEtran}
\bibliography{./bibliography/IEEEabrv,./bibliography/IEEEexample}

\vspace{12pt}

\end{document}